\documentclass{jfm}

\usepackage{graphicx}
\usepackage{natbib}
\usepackage{epstopdf}
\usepackage{amsmath}
\usepackage{esdiff}
\usepackage{subfig}

\ifCUPmtlplainloaded \else
  \checkfont{eurm10}
  \iffontfound
    \IfFileExists{upmath.sty}
      {\typeout{^^JFound AMS Euler Roman fonts on the system,
                   using the 'upmath' package.^^J}%
       \usepackage{upmath}}
      {\typeout{^^JFound AMS Euler Roman fonts on the system, but you
                   dont seem to have the}%
       \typeout{'upmath' package installed. JFM.cls can take advantage
                 of these fonts,^^Jif you use 'upmath' package.^^J}%
      }
  \else
  \fi
\fi

\ifCUPmtlplainloaded \else
  \checkfont{msam10}
  \iffontfound
    \IfFileExists{amssymb.sty}
      {\typeout{^^JFound AMS Symbol fonts on the system, using the
                'amssymb' package.^^J}%
       \usepackage{amssymb}%
       \let\le=\leqslant  
       \let\ge=\geqslant  
      }{}
  \fi
\fi

\ifCUPmtlplainloaded \else
  \IfFileExists{amsbsy.sty}
    {\typeout{^^JFound the 'amsbsy' package on the system, using it.^^J}%
     \usepackage{amsbsy}}
    {}
\fi

\newsavebox{\astrutbox}
\sbox{\astrutbox}{\rule[-5pt]{0pt}{20pt}}

\title{Axial Creeping Flow in the Gap between a Rigid Cylinder and a Concentric Elastic Tube}
\shorttitle{Creeping Flow in the Gap Between a Rigid Cylinder and an Elastic Tube}
\author[S. B. Elbaz and A. D. Gat]{S. B. Elbaz and A. D. Gat}
\affiliation{Faculty of Mechanical Engineering, Technion - Israel Institute of Technology, Haifa 32000, Israel}

\pubyear{2015} \volume{999} \pagerange{999--9999}
\date{2015}

\begin{document}

\maketitle

\begin{abstract}
We examine transient axial creeping flow in the annular gap between a rigid cylinder and a concentric elastic tube. The gap is initially filled with a thin fluid layer. The study focuses on viscous-elastic time-scales for which the rate of solid deformation is of the same order-of-magnitude as the velocity of the fluid. We employ an elastic shell model and the lubrication approximation to obtain a forced nonlinear diffusion equation governing the viscous-elastic interaction. In the case of an advancing liquid front into a configuration with a negligible film layer (compared with the radial deformation of the elastic tube), the governing equation degenerates into a forced porous medium equation, for which several closed-form solutions are presented. In the case where the initial film layer is non-negligible, self-similarity is used to devise propagation laws for a pressure driven liquid front. When advancing external forces are applied on the tube, the formation of dipole structures is shown to dominate the initial stages of the induced flow and deformation regimes. These are variants of the dipole solution of the porous medium equation. Finally, since the rate of pressure propagation decreases with the height of the liquid film, we show that isolated moving deformation patterns can be created and superimposed to generate a moving wave-like deformation field. The presented interaction between viscosity and elasticity may be applied to fields such as soft-robotics and micro-scale or larger swimmers by allowing for the time-dependent control of an axisymmetric compliant boundary.
\end{abstract}

\section{Introduction}
We examine the effect of elasticity on transient axial creeping flow in the annular gap between a rigid cylinder and a concentric elastic tube. Specifically, we aim to analyze nonlinear viscous-elastic dynamics where the elastic deformation significantly changes the boundaries of the flow. The gap between the cylinder and tube is assumed small compared to the radius and length of the cylinder. The flow-field is modeled by applying the lubrication approximation, while the external elastic tube is modeled by thin shell theory. We focus on viscous-elastic time-scales for which the rate of solid deformation is of the same order-of-magnitude as the velocity of the flow.  

A similar configuration was examined by \cite{paidoussis1998fluid} who focused on high Reynolds numbers. The interaction of elastic tubes and thin viscous flows, at the limit of low Reynolds numbers, has been studied by \cite{halpern1992fluid} and \cite{white2005three} who, among others, analyzed the dynamics of a liquid film coating the inner surface of an elastic tube in the context of flows in small airways in the lungs. Other studies involving a thin creeping flow between a rigid surface and an elastic surface focused on planar configurations. These include \cite{chauhan2002settling}, who modeled the dynamics of a contact lens during blinking as a thin viscous film contained between an elastic shell and a flat rigid surface. \cite{pihler2012suppression} and \cite{al2013two} studied the effect of elasticity on the onset of Taylor-Saffman fingering instability in Hele-Shaw cells. \cite{pihler2014interaction} related the patterns of viscous fingering to patterns of wrinkling in an elastic Hele-Shaw cell. \cite{trinh2014pinned} and \cite{trinh2014elastic} studied rigid and elastic plates, either pinned or free floating, moving over a viscous film laying on a flat rigid surface. \cite{carlson2015elastohydrodynamics} studied the deformation and flow-field created by a propagating adhesion front attaching an elastic sheet to a rigid surface, where the gap between the elastic sheet and rigid surface is filled with a viscous fluid. 

For the case in which there is initially a vanishingly small gap between the rigid cylinder and the elastic tube, the examined problem can be viewed as a cylindrical version of the peeling problem \citep{mcewan1966peeling,hosoi2004peeling,lister2013viscous}. In this regard the peeling formation of the current study bears mathematical proximity to a viscous gravity current \citep[e.g.,][]{buckmaster1977viscous,huppert1982propagation,momoniat2006axisymmetric}. Gravity currents have also been studied in geophysical contexts when coupled with elastic surfaces \citep{howell2013gravity,hewitt2015elastic,balmforth2015speed}.

To the best of our knowledge, the problem examined in this work has not previously been studied. The suggested configuration may have bearing on models of compliant boundaries \citep{gad2002compliant}, axisymmetric swimmers and soft robotic applications. The structure of this work is as follows: In \S2 the geometry, relevant parameters and physical assumptions are defined. In \S3 elastic shell theory and the lubrication approximation are employed to obtain a forced nonlinear diffusion equation governing the viscous-elastic interaction. \S4 examines an advancing liquid front in the limit of an initial vanishingly small gap (viscous peeling) where the governing equation simplifies to a porous medium equation. In \S5 the self-similar scheme of the porous medium equation is modified to account for a pre-wetted gap between the cylinder and tube. In \S6 the dipole solution of the porous medium equation is related to the pressure and deformation field created in the initial stages of the response to an advancing external pressure acting on the elastic tube. \S7 examines isolated moving deformations for the case of relative axial speed between the rigid cylinder and the elastic tube. \S8 presents concluding remarks.

\section{Problem formulation}
\begin{figure}
\centering
\includegraphics[width=1\textwidth]{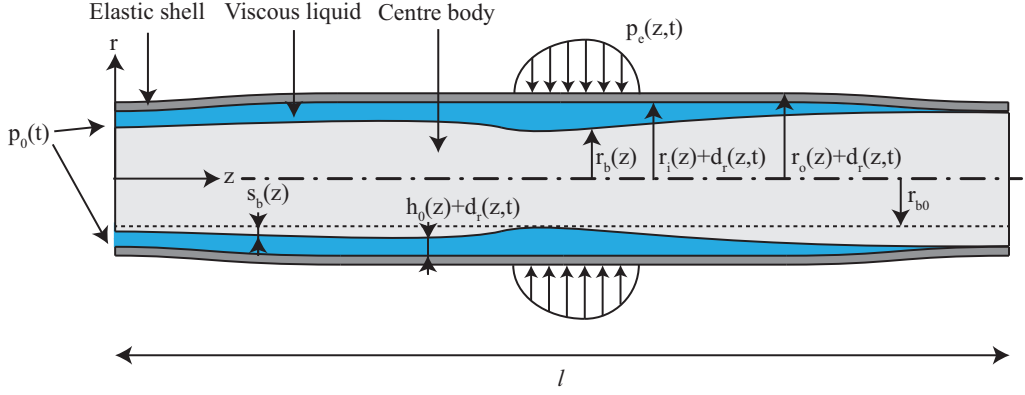}
\caption{Schematic illustration of the elastic tube and annular fluid layer enclosing a weakly varying cross-section rigid centre-body. We define the cylindrical coordinates $(r,z)$. $p_e$ is the external pressure field, and $d_r,d_z$ are radial and axial solid deformations, respectively. $r_b(z)$ is the radius of the centre-body, $r_{b0}$ is the minimal value of $r_b(z)$ and $s_b(z)=r_{b}(z)-r_{b0}$. $r_i(z)+d_r(z,t)$ and $r_o(z)+d_r(z,t)$ are the inner and outer surfaces of the elastic tube, respectively. $h_0(z)+d_r(z,t)$ is the annular gap between the inner centre-body and the elastic tube. $p_0(t)$ is an arbitrary time varying inlet pressure.}
\label{figure_1}
\end{figure}

We study Newtonian, incompressible axial creeping flow in the annular gap between a rigid cylinder and a concentric linearly elastic tube. The gap between the cylinder and tube is assumed small compared to the radius of the cylinder. The flow is driven by time-varying pressures at the inlet or outlet, relative axial speed between the centre-body (cylinder) and elastic tube, as well as an external pressure field acting on the elastic tube. We examine transient dynamics in the viscous-elastic time-scale, where the rate of elastic deformation is of the same order-of-magnitude as the fluid velocity. We assume sufficient height of the liquid filled gap between the tube and centre-body, so that Van-der-Waals forces between the solids can be neglected. While \S4-7 deal with configurations with constant radii of the centre-body and tube when at rest, in \S2-3 we allow for small axial variations of the configuration (see Fig. \ref{figure_1}).

%\subsection{Nomenclature}
Hereafter, normalized variables are denoted by uppercase letters and characteristic parameters are denoted by lowercase letters with asterisks (e.g., if $a$ is a dimensional variable, $a^*$ is the characteristic value of $a$ and $A=a/a^*$ is the corresponding normalized variable).

The relevant variables and parameters are time $t$, axial coordinate $z$, radial coordinate $r$, axial liquid speed $u_z$, radial liquid speed $u_r$, liquid pressure $p$, liquid viscosity $\mu$, liquid density $\rho$, solid axial deformation $d_z$, solid radial deformation $d_r$, solid strain $e_{ij}$ and stress $\sigma_{ij}$ (acting on the plane normal to coordinate $i$ and in the direction of coordinate $j$), solid Young's modulus $E$, solid Poisson's ratio $\nu$, external shell (tube) inner contour $r_i(z)$ and outer contour $r_o(z)$ (see Fig. \ref{figure_1}), external shell midsection, $r_m(z)=(r_i(z)+r_o(z))/2$, centre-body contour radius $r_b (z)$ and the length of the configuration in the $z$ direction, $l$.

The gap (at rest) between the inner body and tube is defined as $h_0(z)=r_i(z)-r_b(z)$. We define the auxiliary radial coordinate, $s=r-r_b(z)$. The base radius of the inner-body is defined by $r_{b0}=\min⁡(r_b (z))$ and the corresponding function describing the centre-body radius is $s_b (z)=r_b (z)-r_{b0}$. We define the slenderness ratio of the configuration as $\epsilon=r_{b0}/l$. The ratio of characteristic fluid layer thickness at rest to characteristic radial deformation is defined by $ {\lambda _h} = {h_0^*}/{d_r^*}$ and the ratio of mean centre-body depth in excess of the base radius to characteristic radial deformation is defined by $\lambda_b = {s_b^*}/{d_r^*}$. 

We define normalized coordinates, $(R,Z)=(r/r_{b0},z/l)$, normalized radial and axial deflections, $(D_{r},D_z)=(d_{r}/d_r^*,d_{z}/d_z^*)$,
normalized liquid pressure, $P=p/p^*$ and external domain pressure, $P_e=p_e/p^*$, where $p^*$ is the characteristic pressure, and normalized liquid velocity	$(U_r,U_z)=(u_r/u_r^*,u_z/u_z^*)$, with corresponding characteristic values $u_r^*$ and $u_z^*$.

%\subsection{Small parameters}
Small parameters in the analysis include the ratio between the change in centre-body cross-section and the length of the geometry in the $z$ direction,
\begin{equation}
	\frac{dr_b}{dz} \sim \frac{s_b^*}{l} \ll 1\,,
	\label{contour_requirement}
\end{equation}
and the ratio between the characteristic radial deformation $d_r^*$ and the length of the geometry in the $z$ direction, i.e., the fluidic aspect ratio,
\begin{equation}
	\frac{{d_r^*}}{l} = {\varepsilon _1} \ll 1\,.
\end{equation}
We utilize a thin elastic shell approximation and thus require the ratio of wall thickness $r_o-r_i$ to inner tube radius $r_i$ to be small,
\begin{equation}
	\frac{{{r_o} - {r_i}}}{{{r_i}}} = {\varepsilon _2} \ll 1\,,
\end{equation}
as well as small elastic deformations, expressed by the ratios,
\begin{equation}
	\frac{{d_r^*}}{{{r_i}}}\sim\frac{{d_z^*}}{l} = {\varepsilon _3} \ll 1\,.
\end{equation}
Our analysis also assumes negligible inertia, represented by the condition,
\begin{equation}
Re_r=	\frac{d_r^*}{l}\frac{\rho u^*_z d_r^*}{\mu}\ll 1\,.
\end{equation}
A more detailed discussion of the condition of negligible inertia is presented in \S3.4.

\section{Obtaining the leading order governing equation of the viscous-elastic dynamics}
In order to obtain the governing viscous-elastic interaction equation, we apply the lubrication approximation for the flow field and the Kirchhoff-Love thin shell approximation for the solid deformation field, under the requirement of similar time-scales of both elastic and fluidic dynamics.

While the elastic shell configuration is similar to our previous work \citep{elbaz2014dynamics}, the characteristic shear stress $\sigma_{zr}^*$ applied by the liquid on the solid is $O(\varepsilon_3)$ smaller compared with the characteristic shear obtained in \cite{elbaz2014dynamics}, (this is readily seen from order-of-magnitude analysis of an integral force balance  $p^*d_r^*2\pi r_i\sim \sigma_{zr}^* l 2\pi r_i$). We thus redevelop the thin shell approximation herein.

\subsection{The elastic problem}
The deformation field of axisymmetric linearly elastic material with negligible inertia is governed by the momentum equations,
\begin{equation} \label{momentum_elastic}
\frac{\partial}{\partial r}\left(r\sigma_{rr}\right)+\frac{\partial}{\partial z}\left(r\sigma_{zr}\right)-\sigma_{\theta\theta}=0\,,\quad
%\end{equation}
%\begin{equation} \label{z_momentum_elastic}
\frac{\partial}{\partial r}\left(r\sigma_{zr}\right)+\frac{\partial}{\partial z}\left(r\sigma_{zz}\right)=0\,,
\end{equation}
the strain-displacement relations,
\begin{equation}
\label{eq:solid strain}
e_{rr}=\frac{\partial d_r}{\partial r},\quad
%\end{equation}
%\begin{equation}
e_{\theta\theta}=\frac{d_r}{r},\quad
e_{zz}=\frac{\partial d_z}{\partial z},\quad
e_{zr}=\frac{1}{2}\left(\frac{\partial d_r}{\partial z}+\frac{\partial d_z}{\partial r}\right),
\end{equation}
and Hooke's law, 
\begin{subequations}
\label{eq:solid Hooke}
\begin{equation}	
Ee_{zr}=(1+\nu)\sigma_{zr}\,,\quad
%\end{equation}
%\begin{equation}
Ee_{rr}=\sigma_{rr}-\nu\left(\sigma_{zz}+\sigma_{\theta\theta}\right),\quad
Ee_{\theta\theta}=\sigma_{\theta\theta}-\nu\left(\sigma_{rr}+\sigma_{zz}\right),
\end{equation}
\begin{equation}
%\end{equation}
%\begin{equation}
Ee_{zz}=\sigma_{zz}-\nu\left(\sigma_{rr}+\sigma_{\theta\theta}\right).
\end{equation}
\end{subequations}
The boundary conditions for the stress applied by the liquid at $r=r_i+d_r$ are 
\begin{subequations}
\label{eq:solid BCs}
\begin{equation}\label{sigma_rr_bc}
\sigma_{rr}(r=r_i+d_r)=-p+2\mu \frac{\partial u_r}{\partial r}\,, \quad \sigma_{zr}(r=r_i+d_r)=-\mu \left(\frac{\partial u_z}{\partial r}+ \frac{\partial u_r}{\partial z}\right),
\end{equation}
and at $r=r_o+d_r$ the externally applied stress is
\begin{equation}
\sigma_{rr}(r=r_o+d_r)=-p_e\,,\quad \sigma_{zr}(r=r_o+d_r)=0\,.
\end{equation}
\end{subequations}

Following elastic shell theory \citep{Mollmann.1981}, we define stress resultants for the forces $n_{ij}$ and moments $m_{ij}$ as
\begin{equation}\label{resul}
n_{ij}=\int_{r_i}^{r_o}{\sigma_{ij}dr}\,,\quad m_{ij}=\int_{r_i}^{r_o}{\sigma_{ij}rdr}\,.
\end{equation}
The curvature of the external shell $d{r_m}/dz$ will introduce only negligible terms, of order $O(\varepsilon_1)$, in the stress resultants (\ref{resul}). However, centre-body curvature $d{r_b}/dz$ will have leading order effects in the fluidic analysis (\S3.2). We multiply the axial momentum equation by $r-r_m$ and integrate over $r$ to obtain a resultant form. We then differentiate with respect to $z$ in order to relate to the radial momentum equation (\ref{momentum_elastic}). This yields
\begin{equation}
\frac{{\partial {n_{zr}}}}{{\partial z}} - \frac{{{\partial ^2}{m_{zz}}}}{{\partial {z^2}}} + {r_i}\frac{{\partial {\sigma _{zr}}\left( {{r_i}} \right)}}{{\partial z}} = O(\varepsilon_2)\,.
\end{equation}
Substituting into the resultant form of the radial momentum equation, 
\begin{equation}
{r_i}p - {r_o}{p_e} + {r_i}\left( {\frac{{{\partial ^2}{m_{zz}}}}{{\partial {z^2}}} - {r_i}\frac{{\partial {\sigma _{zr}}\left( {{r_i}} \right)}}{{\partial z}}} \right) - {n_{\theta \theta }} = O(\varepsilon_2)\,.
\label{res_r_momentum_elastic}
\end{equation}
Order-of-magnitude analysis of the axial momentum equation in (\ref{momentum_elastic}) yields a relation between characteristic values,
\begin{equation}
 n_{zz}^*={d_r^* p^*},\,\,\, \sigma _{zz}^* = \frac{{{p^*}}}{{{\varepsilon _2}}}{\varepsilon _3}\,,
\end{equation}
and order-of-magnitude analysis of (\ref{res_r_momentum_elastic}) yields
\begin{equation}
n_{\theta \theta }^* = {r_{b0}}{p^*},\,\,\,\sigma _{\theta \theta }^* = \frac{{{p^*}}}{{{\varepsilon _2}}}\,.
\end{equation}
The shear and normal stress characteristic values, $\sigma_{rr}^*=p^*$ and $\sigma_{zr}^*={\varepsilon_1 p^*}$ are obtained from the fluidic problem (see \S3.2).

Taking note that $m_{zz}^* = {r_{b0}}n_{zz}^*$, normalizing (\ref{res_r_momentum_elastic}) yields in leading order,
\begin{equation}
{N_{\theta \theta }}=P - {P_e} + O(\varepsilon_1 \epsilon,\varepsilon_2)\,,
\label{N_tt_P}
\end{equation}
and normalizing Hooke's law (\ref{eq:solid Hooke}) according to the above characteristic values yields a reduced form of Love's first approximation \citep{Love.1888}, 
\begin{equation}
{{\rm{\Sigma }}_{\theta \theta }} \sim - \frac{{E{\varepsilon _2}}}{{{p^*}\nu }}{e_{zz}}\sim \frac{{E{\varepsilon _2}}}{{{p^*}}}{e_{\theta \theta }}\,,
\label{loves_first}
\end{equation}
for which orders of $O\left( {{\varepsilon _2},{\varepsilon _3}} \right)$ are neglected and $\Sigma_{\theta \theta}=\sigma_{\theta \theta}/(p^*/\varepsilon_2)$. We apply \citep{Dugdale.1971} the Kirchhoff hypothesis and describe the displacement field in terms of the radial $\bar{d}_r$ and axial $\bar{d}_z$ displacements of the midsection, denoted by overbars,
\begin{equation}
\bar{d}_z=d_z+(r-r_m)\frac{\partial d_r}{\partial z}\,,\,\,\, \bar{d}_r=d_r\,,
\end{equation}
and thus we can represent the strain as a function of the deformation by
\begin{equation}
e_{zz}=\frac{\partial \bar{d}_z}{\partial z}-(r-r_m)\frac{\partial^2 \bar{d}_r}{\partial z^2}\,,\quad e_{\theta\theta}=\frac{\bar{d}_r}{r}\,.
\end{equation}
Integrating (\ref{loves_first}) over $r$ into resultant form and substituting $N_{\theta \theta}$ from (\ref{N_tt_P})  provides the leading order fluidic pressure to elastic deformations relation,
\begin{equation}
\bar D_r=P - {P_e}\,,\quad \bar D_z-\bar D_z(Z=0)=-\nu\int_0^Z{\left(P - {P_e}\right)dZ}\,,
\label{elastic_press_defrom}
\end{equation}
as well as a relation between the characteristic radial deformation and characteristic pressure,
\begin{equation}
\frac{p^*}{E\varepsilon_2}=\frac{d_r^*}{r_{b0}}=\varepsilon_3\ll1\,.
\label{eps3_cond}
\end{equation}
Equation (\ref{eps3_cond}) provides limitation on the maximal allowed pressure for which the assumption of small deformations is valid.

\subsection{The fluidic problem}
We assume an axisymmetric incompressible Newtonian fluid, governed by the momentum equations,
\begin{equation}
\rho \left( {\frac{{\partial {u_r}}}{{\partial t}} + {u_r}\frac{{\partial {u_r}}}{{\partial r}} + {u_z}\frac{{\partial {u_r}}}{{\partial z}}} \right) =  - \frac{{\partial p}}{{\partial r}} + \mu \left[ {\frac{1}{r}\frac{\partial }{{\partial r}}\left( {r\frac{{\partial {u_r}}}{{\partial r}}} \right) + \frac{{{\partial ^2}{u_r}}}{{\partial {z^2}}} - \frac{{{u_r}}}{{{r^2}}}} \right]\,,
\label{fluid_momentum_r}
\end{equation}
\begin{equation}
\rho \left( {\frac{{\partial {u_z}}}{{\partial t}} + {u_r}\frac{{\partial {u_z}}}{{\partial r}} + {u_z}\frac{{\partial {u_z}}}{{\partial z}}} \right) =  - \frac{{\partial p}}{{\partial z}} + \mu \left[ {\frac{1}{r}\frac{\partial }{{\partial r}}\left( {r\frac{{\partial {u_z}}}{{\partial r}}} \right) + \frac{{{\partial ^2}{u_z}}}{{\partial {z^2}}}} \right] + \rho g\,,
\label{fluid_momentum_z}
\end{equation}
and conservation of mass,
\begin{equation}
\frac{1}{r}\frac{\partial }{{\partial r}}\left( {r{u_r}} \right) + \frac{{\partial {u_z}}}{{\partial z}} = 0\,.
\label{cont_eq}
\end{equation}
The relevant boundary conditions are no-slip and no-penetration at $r=r_i+d_r$,
\begin{equation}
		u_r(r=r_b+h_0+d_r)=\frac{\partial d_r}{\partial t}\,,\,\,\,u_z(r=r_b+h_0+d_r))=\frac{		\partial d_z}{\partial t}\,,
\label{bc_u_fluid_shell}
\end{equation}
and at the centre-body boundary, $r=r_b$,
\begin{equation}
		u_r(r=r_b)=0\,,\,\,\,u_z(r=r_b)=u_0\,.
\label{bc_u_fluid_cb}
\end{equation}
Pressure at the inlet and outlet may be governed by
\begin{equation}
	p(z=0)=p_0(t)\,,\quad p(z=l)=p_l(t)\,,
	\label{bc_u_fluid_p}
\end{equation}
where $p_0(t)$ and $p_l(t)$ are arbitrary functions of time. 
%A corresponding pressure function for the right ($z=l$) inlet may also be defined, however we shall focus on single inlet actuation for which the length-scale of interaction is smaller than the overall length $l$. 
%A boundary value problem for which dual boundary effects can be further examined numerically will be formulated in section (...).
We normalize the coordinate $s=r-r_b(z)$, defined in the gap $0 \le  s \le {h_0}\left( z \right) + {d_r}\left( {z,t} \right)$, by $d_r^*$,
%\end{equation}
%mention the curvature assumptions here how u_r=u_s curvilinear NS- cylindrical N-S
%The truncation is required since the elastic shell cannot contract beneath the solid surface. which takes the normalized form,
\begin{equation}\label{S_define}
S = \frac{{s}}{{d_r^*}}\,,\,\,\,\,\,0 \le S \le {{\rm{\lambda }}_h}{H_0}(Z) + {D_r}(Z,T)\,,
\end{equation}
where $H_0=h_0/h_0^*$. Based on (\ref{S_define}) we can relate the normalized radial coordinate $R$ to $S$,
\begin{equation}
R = 1 + {\varepsilon _3}\left( { S + {\lambda _b}{S_b}\left( { Z} \right)} \right),
\label{R_Z_to_Sbar_Zbar_trans}
\end{equation}
where ${S_b}\left( Z \right) = {s_b}\left( z \right)/s_b^*$. 
%It follows that
%\begin{equation}
%\frac{\partial }{{\partial R}} = \frac{1}{{{\varepsilon _3}}}\frac{\partial }{{\partial \bar S}},\,\,\,\,\,\,\,\,\,\frac{\partial }{{\partial Z}} =  - {\lambda _b}S_b^{'}\frac{\partial }{{\partial \bar S}} + \frac{\partial }{{\partial \bar Z}}\,.
%\label{R_Z_to_Sbar_Zbar_diff}
%\end{equation}
Order-of-magnitude analysis of (\ref{cont_eq}) yields
\begin{equation}
\frac{{u_r^*}}{{u_z^*}}\sim\frac{{d_r^*}}{l} = {\varepsilon _1}\,,\,\,\,\,\,u_z^* = \frac{{{\varepsilon _1}d_r^*{p^*}}}{\mu }\,.
\label{char_speed_ratio}
\end{equation}
We require $u_0 \sim u_z^*$ so that the viscous stresses resulting from centre-body motion scale as $\mu {u_0}/d_r^*$.
We focus on negligible gravity, $G=\rho gl/{p^*} \ll 1$, and define a reduced Reynolds number, $Re = {{\rho u_z^* d_r^*}}/{\mu }$, and Womersley number, $\alpha^2={\rho d{{_r^*}^2}}/{{\mu {t^*}}}$. Applying transformation (\ref{R_Z_to_Sbar_Zbar_trans}) to (\ref{fluid_momentum_r})-(\ref{cont_eq})  and employing (\ref{char_speed_ratio}) yields the following reduced system for the fluidic domain,
\begin{subequations}\label{reduced_eq}
\begin{equation}
\frac{{\partial P}}{{\partial Z}} = \frac{{{\partial ^2}{U_z}}}{{\partial {{ S}^2}}} + O\left( {\alpha^2,{\varepsilon _3},{\varepsilon _1}Re,\varepsilon _1^2,{\lambda _b}\frac{{\partial P}}{{\partial  S}}},G\right),
\label{Reduced_NS_z}
\end{equation}

\begin{equation}
\frac{{\partial P}}{{\partial S}} = O\left( {\alpha^2,\varepsilon _1^3Re,\varepsilon _1^2} \right),
\label{Reduced_NS_r}
\end{equation}

\begin{equation}
\frac{{\partial {U_r}}}{{\partial S}} - {\lambda _b}\frac{{d{S_b}}}{{dZ}}\frac{{\partial {U_z}}}{{\partial S}} + \frac{{\partial {U_z}}}{{\partial Z}} = O\left( {{\varepsilon _3}} \right).
\label{Reduced_cont_eq}
\end{equation}
\end{subequations}
Equations (\ref{reduced_eq}) are valid for ${\lambda _h},{\lambda _b} \sim 1$ and ${\lambda _h},{\lambda _b} \ll 1$. $t^*$ represents the time-scale of the viscous-elastic interaction and is to be obtained by relation to the elastic problem. %This system will provide the foundation for the analysis of the nonlinear interaction of the tube and fluid layer, typically addressing small values of ${\lambda _h}$. 

\subsubsection{Nonlinear problem formulation}
The boundary conditions (\ref{bc_u_fluid_shell}), (\ref{bc_u_fluid_cb}) take the normalized form,
\begin{subequations}
\begin{equation}
{U_r}\left( { S = {{{\lambda _h}{H_0} + {D_r}} }} \right) = \frac{{d_r^*}}{{{t^*}u_r^*}}\frac{{\partial {D_r}}}{{\partial T}}\,,\,\,\,\,\,{U_z}\left( { S = {{{\lambda _h}{H_0} + {D_r}} }} \right) = \frac{{d_z^*}}{{{t^*}u_z^*}}\frac{{\partial {D_z}}}{{\partial T}}\,,
\label{bc_norm_u_fluid_shell}
\end{equation}
and
\begin{equation}
{U_r}\left( { S = 0} \right) = 0\,,\,\,\,\,\,\,{U_z}\left( { S = 0} \right) = {U_0}\,,
\label{bc_norm_u_fluid_cb}
\end{equation}
\end{subequations}
with $U_0=u_0/u_z^*$ and $T=t/t^*$. We solve (\ref{Reduced_NS_z}), (\ref{Reduced_NS_r}) imposing conditions (\ref{bc_norm_u_fluid_shell}), (\ref{bc_norm_u_fluid_cb}). The resulting leading order axial speed $U_z$ reads
\begin{equation}
{U_z}\sim\frac{1}{2}\frac{{\partial P}}{{\partial Z}}\left[ {{{S}^2} - {{\left( {{\lambda _h}{H_0} + {D_r}} \right)} } S} \right] + \frac{{ S}}{{{{\left( {{\lambda _h}{H_0} + {D_r}} \right)} }}}\frac{{d_z^*}}{{{t^*}u_z^*}}\frac{{\partial {D_z}}}{{\partial T}} + {U_0}\left[ {1 - \frac{{ S}}{{{{\left( {{\lambda _h}{H_0} + {D_r}} \right)} }}}} \right]\,,
\label{axial_vel_profile}
\end{equation}
and is defined for ${{\lambda _h}{H_0} + {D_r}} > 0$. Addressing mass conservation (\ref{Reduced_cont_eq}), we install (\ref{axial_vel_profile}) and integrate with respect to $S$ across the film layer to produce a reduced Reynolds equation relating fluidic pressure to elastic deformations and curvature,
%alignment can be fixed using amsmath
\begin{eqnarray}
&&\frac{{d_r^*}}{{{t^*}u_r^*}}\frac{{\partial {D_r}}}{{\partial T}} - \frac{1}{{12}}\frac{\partial }{{\partial Z}}\left\{ {\frac{{\partial P}}{{\partial Z}}{{\left( {{\lambda _h}{H_0} + {D_r}} \right)}^3 }} \right\}
+{U_0}\left[ {\frac{1}{2}\frac{\partial }{{\partial Z}}{{\left( {{\lambda _h}{H_0} + {D_r}} \right)} } + {\lambda _b}\frac{{d{S_b}}}{{dZ}}} \right] \nonumber 
\\
&&+\frac{1}{2}\frac{{{d_z}^*}}{{{t^*}u_z^*}}\left[ {{{\left( {{\lambda _h}{H_0} + {D_r}} \right)} }\frac{{{\partial ^2}{D_z}}}{{\partial Z\partial T}} - \left[ {\frac{\partial }{{\partial Z}}{{\left( {{\lambda _h}{H_0} + {D_r}} \right)} } + 2{\lambda _b}\frac{{d{S_b}}}{{dZ}}} \right]\frac{{\partial {D_z}}}{{\partial T}}} \right]\sim0\,.
\label{Fluid_Rey_eq}
\end{eqnarray}
Equation (\ref{Fluid_Rey_eq}) is valid for ${{\lambda _h}{H_0} + {D_r}} > 0$ and must be zeroed for values of $(Z,T)$ that render ${{\lambda _h}{H_0} + {D_r}} \le 0$, since fluid pressure cannot contract the tube beneath the centre-body surface. We are now in a position to evaluate the characteristic coefficients and hence the time-scale $t^*$. By dominant balance a consistent result is achieved by setting ${t^*} = d_r^*/u_r^*$. The residual axial speed terms, coefficients of $\partial {D_z}/\partial T$, are of order $O\left( {{\varepsilon _3}} \right)$. Thus (\ref{Fluid_Rey_eq}) reduces to
\begin{equation}
\frac{{\partial {D_r}}}{{\partial T}} - \frac{1}{{12}}\frac{\partial }{{\partial Z}}\left\{ {\frac{{\partial P}}{{\partial Z}}{{\left( {{\lambda _h}{H_0} + {D_r}} \right)}}^3} \right\}+{U_0}\left[ \frac{1}{2}\frac{{\partial }}{{\partial Z}}\left({\lambda_h H_0}+{D_r}\right) + {\lambda _b}\frac{{d{S_b}}}{{dZ}} \right]\sim0\,.
\label{Reduced_Fluid_Rey_eq}
\end{equation}
%$s_b^* = \left( {1/l} \right)\mathop \smallint \limits_0^l {s_b}\left( z \right)dz$'

\subsubsection{Linear problem formulation}
In the case of large $\lambda_h=h_0^*/d_r^*\gg1$ the characteristic gap should be defined by $h_0^*$ and not $d_r^*$. This requires renormalization of the lubrication approximation attained in (\ref{Reduced_NS_z})-(\ref{Reduced_cont_eq}). The renormalized parameters associated with the linear problem, denoted by the tilde symbol, are given by
In the case of large $\lambda_h=h_0^*/d_r^*\gg1$ the characteristic gap should be defined by $h_0^*$ and not $d_r^*$. This requires renormalization of the lubrication approximation attained in (\ref{Reduced_NS_z})-(\ref{Reduced_cont_eq}). The renormalized parameters associated with the linear problem, denoted by the tilde symbol, are given by
\begin{equation}
\tilde \varepsilon _1 = \lambda_h \varepsilon_1,\,\,\, \tilde \varepsilon _3 = \lambda_h \varepsilon_3,\,\,\, \tilde \lambda _b = \lambda _h ^{-1} \lambda_b,\,\,\, \widetilde Re = \lambda _h Re,\,\,\, \tilde \alpha = \lambda _h \alpha\,. 
\label{wiggle_params}
\end{equation}
We renormalize the local film coordinate and domain, 
\begin{equation}
\widetilde {S} = \lambda_h^{-1} S\,,\,\,\,\,\,0 \le \widetilde {S} \le {H_0}(Z) + {{\rm{\lambda }}_h^{-1}}{D_r}(Z,T)\,,
\end{equation}
and reapply boundary conditions (\ref{bc_norm_u_fluid_shell})-(\ref{bc_norm_u_fluid_cb}) in $\widetilde {S}$. The characteristic radial and axial velocities now scale as ${u_r^*}/{u_z^*}\sim {\tilde \varepsilon _1}$ and $u_z^* ={\tilde \varepsilon _1 h_0^*p^*}/{\mu}$. The reduced equations of motion for the linear case may be rewritten in tilde form,
\begin{subequations}
\begin{equation}
\frac{{\partial P}}{{\partial Z}} = \frac{{{\partial ^2}{U_z}}}{{\partial {{\widetilde S}^2}}} + O\left( {\tilde \alpha^2,{\tilde \varepsilon _3},{\tilde \varepsilon _1} \widetilde Re,\tilde \varepsilon _1^2,{\tilde \lambda _b}\frac{{\partial P}}{{\partial \widetilde S}}},G\right),
\label{Reduced_NS_z_wig}
\end{equation}

\begin{equation}
\frac{{\partial P}}{{\partial \widetilde S}} = O\left( {\tilde \alpha^2,\tilde\varepsilon _1^3 \widetilde Re, \tilde \varepsilon _1^2} \right),
\label{Reduced_NS_r_wig}
\end{equation}

\begin{equation}
\frac{{\partial {U_r}}}{{\partial \widetilde S}} - {\tilde \lambda _b}\frac{{d{S_b}}}{{dZ}}\frac{{\partial {U_z}}}{{\partial \widetilde S}} + \frac{{\partial {U_z}}}{{\partial Z}} = O\left( {{\tilde \varepsilon _3}} \right).
\label{Reduced_cont_eq_wig}
\end{equation}
\end{subequations}
Following a similar procedure to that of \S 3.2.1 we derive a linearized reduced Reynolds equation analogous to (\ref{Reduced_Fluid_Rey_eq}),
\begin{equation}
\frac{{\partial {D_r}}}{{\partial T}} - \frac{1}{{12}}\frac{\partial }{{\partial Z}}\left\{ {\frac{{\partial P}}{{\partial Z}}{{\left( {{H_0} + {{{\lambda }}_h^{-1}}{D_r}} \right)}}^3} \right\} + {U_0}\left[ \frac{1}{2}\frac{{\partial }}{{\partial Z}}\left({H_0}+{{{\lambda }}_h^{-1}}{D_r}\right) + {\tilde \lambda _b}\frac{{d{S_b}}}{{dZ}} \right]\sim0\,.
\label{Lin_Reduced_Fluid_Rey_eq}
\end{equation}
%Since linear dynamics are not the focus of this study, solutions of (\ref{Lin_Reduced_Fluid_Rey_eq}) will not be presented in this work.

\subsection{The fluidic-elastic problem}
We substitute (\ref{elastic_press_defrom}) into (\ref{Reduced_Fluid_Rey_eq}) dropping the midsection overbars and derive an initial boundary value problem for the nonlinear viscous-elastic interaction (for the corresponding dimensional equation see equation \ref{NL_IBVL_P_dim} of Appendix \ref{AppA}),
\begin{equation}
\frac{{\partial {(P-P_e)}}}{{\partial T}} - \frac{1}{{12}}\frac{\partial }{{\partial Z}}\left\{ {\frac{{\partial P}}{{\partial Z}}{{\left( {{\lambda _h}{H_0} + {P-P_e}} \right)}}^3} \right\}+{U_0}\left[ \frac{1}{2}\frac{{\partial }}{{\partial Z}}\left({\lambda_h H_0}+{P-P_e}\right) + {\lambda _b}\frac{{d{S_b}}}{{dZ}} \right]\sim0\,,
\label{NL_IBVL_P}
\end{equation}
with corresponding centre-body surface condition rewritten in pressure form,
\begin{equation}
P>P_e-\lambda_hH_0\,.
\label{surface_cond_pres}
\end{equation}
Equations (\ref{NL_IBVL_P})-(\ref{surface_cond_pres}) may be solved with an initial condition $P(Z,0)=P_i(Z)$ and boundary conditions of type (\ref{bc_u_fluid_p}) in normalized form $P(0,T)=P_0(T),P(1,T)=P_1(T)$. The characteristic time-scale $t^*$ of the nonlinear viscous-elastic interaction is evaluated via relations (\ref{eps3_cond}) and (\ref{char_speed_ratio}),
\begin{equation}
t^*=\frac{(E\varepsilon_2)^2\mu}{{p^*}^3\epsilon^2}\,.
\label{NL_time_scale}
\end{equation}
The dependence in characteristic pressure stems from the pressure dependent diffusion coefficient of (\ref{NL_IBVL_P}). 

As per the linear case, all elastic analysis results of \S 3.1 hold with substituted tilde parameters of (\ref{wiggle_params}) where applicable. Therefore an analogous substitution of (\ref{elastic_press_defrom}) into (\ref{Lin_Reduced_Fluid_Rey_eq}) will yield the corresponding linear initial boundary value problem for inlet pressure driven viscous-elastic perturbations of the external shell,
\begin{equation}
\begin{split}
\frac{{\partial {(P-P_e)}}}{{\partial T}} - \frac{1}{{12}}\frac{\partial }{{\partial Z}}\left\{ {\frac{{\partial P}}{{\partial Z}}{{\left( {{H_0} + {{{\lambda }}_h^{-1}}{(P-P_e)}} \right)}}^3} \right\}\\ + {U_0}\left[ \frac{1}{2}\frac{{\partial }}{{\partial Z}}\left({H_0}+{{{\lambda }}_h^{-1}}{(P-P_e)}\right) + {\tilde \lambda _b}\frac{{d{S_b}}}{{dZ}} \right]\sim0\,.
\label{LIN_IBVL_P}
\end{split}
\end{equation}
With corresponding initial and boundary conditions. The characteristic time-scale $t^*$ is reevaluated for the linear case, %with the aid of (\ref{LIN_IBVL_P}) and (\ref{eps3_cond}) yielding,
\begin{equation}
t^*=\frac{\mu}{E\tilde \varepsilon _1 ^2 \tilde \varepsilon _3 \varepsilon_2}\,,
\label{LIN_time_scale}
\end{equation}
and shows exclusive dependence in solid-liquid material properties and the geometry of the configuration. A result which is consistent with the linear interaction time-scale of the full cross-section cylindrical shell \citep{elbaz2014dynamics}.
%organize final LIN and NL IBVPs, rescale U_0?

\subsection{Non-linearity and neglected inertia terms}
Equating the non-linear time-scale, (\ref{NL_time_scale}) and linear time-scale, (\ref{LIN_time_scale}) we derive a condition on the characteristic pressure for which the system will exhibit nonlinear behavior,
\begin{equation}
p^* \ge E \lambda_h \varepsilon_2 \varepsilon _3 = E\frac{r_o-r_i}{r_i}\frac{h^*_0}{r_i}\,.
\label{Cnd_NL_P}
\end{equation}
As the pre-wetting layer vanishes, $h_0^*\to 0$, any small perturbation in pressure will propagate nonlinearly, representing a state for which the problem is not linearizable. There is an upper limit on the characteristic pressure beyond which the assumption of negligible inertia (\ref{Reduced_NS_z})-(\ref{Reduced_NS_r}) will no longer be valid, representing the range of validity of the analysis. Substituting (\ref{NL_time_scale}) into (\ref{Reduced_NS_z}), we demand that both $\alpha^2$ and $\varepsilon_1Re$ terms be negligible and derive the condition that
\begin{equation}
p^*\ll \left(\frac{\mu}{\epsilon r_{b0}}\right)^{\frac{2}{5}}\frac{\left(E\varepsilon_2\right)^{\frac{4}{5}}}{\rho^{\frac{1}{5}}}\,.
\label{Cnd_NL_P_inertia}
\end{equation}
It will be shown in \S 5 that the transition from a nonlinear to a linear regime occurs quite rapidly in the intermediate ${\lambda _h}$ range.
%As the studied viscous-elastic interaction is principally pressure driven the above conditions define the order-of-magnitude of the characteristic pressure for a given set of material properties and geometrical scale.
% Can also supplement linear conditions (notes 56)

\section{Closed-form solution of the peeling problem and derivation of propagation laws}
At the limit $\lambda_h \to 0$ while setting $P_e=U_0=0$, (\ref{NL_IBVL_P}) describes cylindrical pressure-driven viscous peeling of the external shell from the centre-body and reduces to the porous medium equation (PME) in fourth power,
\begin{equation}
\frac{{\partial P}}{{\partial T}} - \frac{{{\partial ^2}{P^4}}}{{\partial {Z^2}}}\sim0{\kern 1pt}\,,
\label{PME_P}
\end{equation}
with adjusted time-scale, $T=t/\kappa t^*, \kappa=48$. We transform to self-similar variables,
\begin{equation}
P=T^{-\alpha}f(\eta),\,\,\, \eta=ZT^{\frac{3\alpha-1}{2}}\,,
\label{Self_sim_PME}
\end{equation}
and adopt Huppert's renormalization \citep{huppert1982propagation} in light of the close mathematical analogy to a viscous gravity current propagating under a density gradient,
\begin{equation}
F(\xi)=\eta_{_F}^{-\frac{2}{3}}f(\eta)\,\,\,, \eta=\eta_{_F}\xi\,.
\label{Huppert_Norm}
\end{equation}
$\eta_{_F}$ describes the propagating peeling front for which the solution is supported in the region $Z<Z_{_F}(T)$, prior to boundary interaction $Z_{_F}(T)<1$. We install (\ref{Self_sim_PME}), (\ref{Huppert_Norm}) into (\ref{PME_P}) which then reduces to the following ODE in $F$,
\begin{equation}
({F^4})'' + \left( {\frac{{1 - 3\alpha }}{2}} \right)\xi F' + \alpha F\sim0\,.
\label{Self_sim_ODE}
\end{equation}
The eigenvalue $\alpha$ describes the inlet pressure boundary condition,
\begin{equation}
P(0,T)=T^{-\alpha}\,,
\label{Self_sim_BC_P}
\end{equation}
and thus,
\begin{equation}
F(0)=\eta_{_F}^{-\frac{2}{3}},\,\,\,F(\xi\ge1)=0\,.
\label{Self_sim_BC_F}
\end{equation}
We complete the formulation with integral mass conservation represented in non-dimensional form by
\begin{equation}
\int_0^{{Z_{_F}}(T)} {P(Z,T)dZ}=QT^{\frac{1-5\alpha}{2}},\,\,\,Q=\eta_{_F}^{\frac{5}{3}}\int_0^1 {F(\xi )d\xi }\,. 
\label{Non_dim_int_MC}
\end{equation}
Both $\eta_{_F}$ and $Q$ are functions of the eigenvalue $\alpha$. Contrary to a viscous gravity current where the interface is induced by a given flux input, in which case the flux rate $Q$ may be set constant for all $\alpha$, in the current study the flux rate derives from the inlet pressure.

\begin{figure}
\centering
\includegraphics[width=1.0\textwidth]{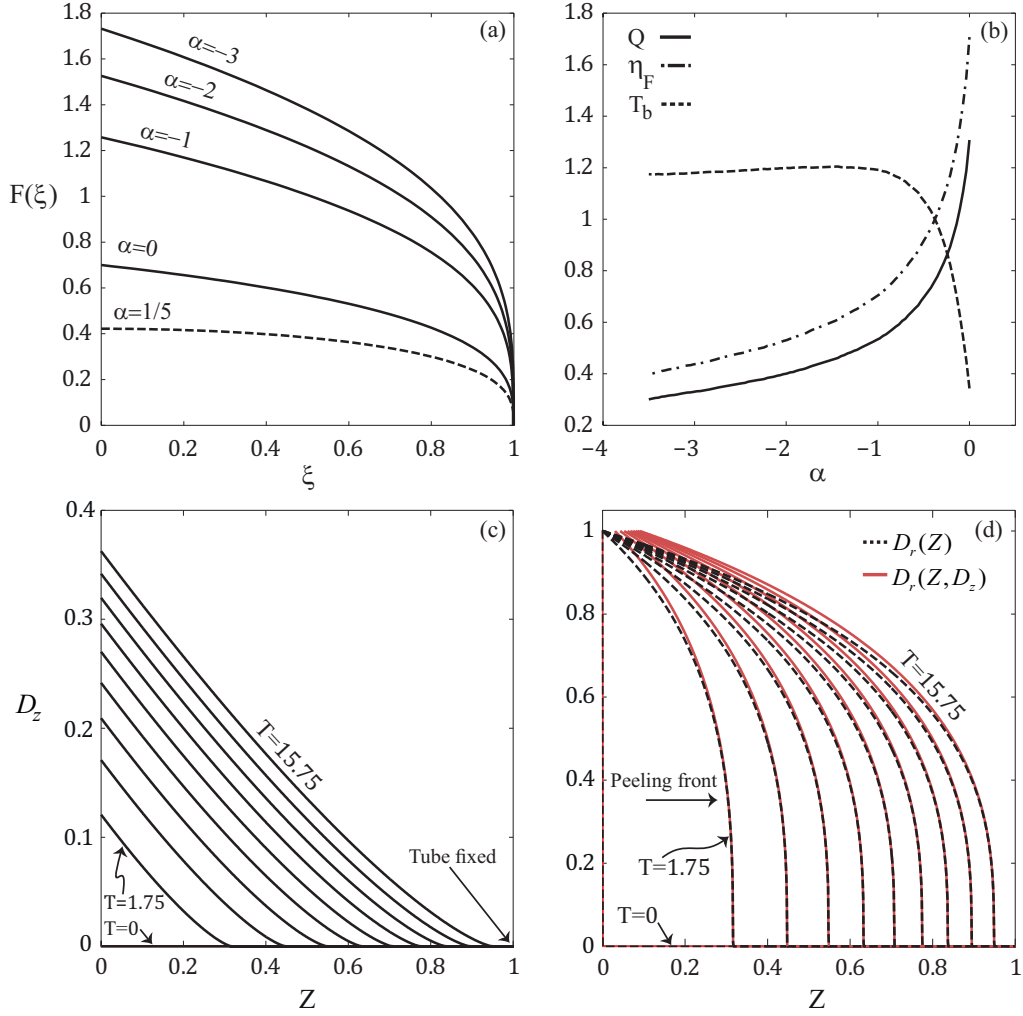}% Here is how to import EPS art
\caption{Self-similar solutions of cylindrical pressure driven viscous peeling, corresponding to $\lambda_h \to0, P_e=U_0=0$ and inlet pressure $P(0,T)= T^{-\alpha}$. Part (a): Pressure profile $F(\xi)$ corresponding to eigenvalues: $\alpha=-1/5$ (analytic) - dashed line, $\alpha=0,-1,-2,-3$ - solid lines (obtained numerically). Part (b): Flux rate $Q(\alpha)$ - solid line, front locus $\eta_{_F}(\alpha)$ - dashed-dotted line, time to boundary $T_b(\alpha)=t_b/\kappa t^*$ - dashed line ($Q,\eta_{_F}$ and $T_b$ obtained numerically). Parts (c,d): Constant boundary pressure viscous peeling ($\alpha=0$), tube fixed at $Z=1$, $D_r(1,T)=D_z(1,T)=0$.  Part (c): Axial deformation $D_z(Z,T)$, progressive snapshots of $\Delta T=1.75$, between $0\le T \le15.75$, shown in the natural time-scale $T=t/t^*$. Part (d): Radial deformation $D_r(Z,T)$ vs. undeformed axial coordinate - dashed black, vs. deformed axial coordinate - solid red, progressive snapshots of $\Delta T=1.75$, between $0\le T \le 15.75$, shown in the natural time-scale $T=t/t^*$, $\varepsilon_3=0.25$, $\nu=0.5$.}
\label{figure_2}
\end{figure}

Eq. (\ref{Self_sim_ODE}) is solved analytically for $\alpha=1/5$ and is a particular case of the source solution of the PME, known as the ZKB solution, as first obtained by \cite{ZK1950PME} and \cite{Barenblatt1952PME}. The underlying boundary and initial conditions are
\begin{equation}
P(Z,0) = Q\delta (Z)\,,\quad \frac{{\partial P(0,T)}}{{\partial Z}}=0\,,
\label{ZKB_IC_BC}
\end{equation}
representing a sudden input of mass, $Q$, into the interface at $T=0$, after which the inlet is sealed. The amplitude decay at the inlet, as the front spreads through the interface, is then given by (\ref{Self_sim_BC_P}). The solution reads
\begin{equation}
P(Z,T) = T^{-\frac{1}{5}}\left[1-\frac{3}{40}Z^2T^{-\frac{2}{5}}\right]_+^{\frac{1}{3}},\,\,\,Q = \frac{{\sqrt {\frac{{10\pi }}{3}} \Gamma \left( {\frac{4}{3}} \right)}}{{\Gamma \left( {\frac{{11}}{6}} \right)}}\,,\,\,\,\eta_{_F}=\sqrt{\frac{40}{3}}\,,
\label{ZKB_SOL}
\end{equation}
where $(s)_{_+}=\max(s,0)$ and $\Gamma$ is Euler's Gamma function. The velocity field corresponding to (\ref{ZKB_SOL}) can be attained via (\ref{axial_vel_profile}) and (\ref{Reduced_cont_eq}). The dimensional solution corresponding to (\ref{ZKB_SOL}) is given by (\ref{ZKB_SOL_dim}). For all other physical values of $\alpha$ ($\alpha<1/5$) (\ref{Self_sim_ODE}), along with conditions (\ref{Self_sim_BC_F}) and (\ref{Non_dim_int_MC}), is solved numerically. In the range $\alpha\le0$ the underlying initial condition is $P(Z,0)=0$. Fig. \ref{figure_2}(a) presents the self-similar pressure profile $F(\xi)$ for $\alpha=1/5$ (\ref{ZKB_SOL}) and for $\alpha=0,-1,-2,-3$ corresponding to inlet signals of type (\ref{Self_sim_BC_P}). Fig. \ref{figure_2}(b) depicts the flux rate $Q$, the front locus $\eta_{_F}$, and the time it takes the front to reach the opposite boundary, denoted $T_b$, as a function of $\alpha$. Fig. \ref{figure_2}(b) is plotted in the range $\alpha\le0$ (the range $0<\alpha\le1/5$ is irrelevant for comparison due to the change in boundary and initial conditions). $T_b$ represents the limit of validity of the self-similar analysis when adapted to a finite domain, it reaches a maxima around $\alpha=-1$ bellow which it decreases moderately. Figs. \ref{figure_2}(c) and (d) depict the resulting deformation regime, $D_z$ and $D_r$, respectively, for the case of constant boundary pressure ($\alpha=0$). The peeling front enters from the left ($Z=0$) while the tube is set stationary at the right end $D_r(1,T)=D_z(1,T)=0$. The deformations are attained via (\ref{elastic_press_defrom}), a strain ratio of $\varepsilon_3=0.25$ and a Poisson's ratio of $\nu=0.5$ were used. As the front propagates the interface the tube shrinks from the free left end. The upper half of the radial deformation profile slightly steepens due to the $D_z$ deformation field.

The propagation laws for an inlet signal of type (\ref{Self_sim_BC_P}), according to self-similarity, are written in terms of the front location $Z_{_F}$, accumulated mass in the interface $M$, and time to boundary $T_b$,
\begin{equation}
{Z_{_F}} = {\eta _{_F}(\alpha)}{T^{\frac{{1 - 3\alpha }}{2}}},\,\,\,M=Q(\alpha){T^{\frac{{1 - 5\alpha }}{2}}},\,\,\,T_b={\eta _{_F}(\alpha)}^{-\frac{2}{1-3\alpha}}\,,
\label{Self_sim_Prop_laws}
\end{equation}
where $T=t/\kappa t^*$ and $\kappa=48$. The corresponding dimensional propagation laws are given in (\ref{eq:dim prop laws}). The spread rate of a disturbance in an unperturbed medium in the case of a full cross-section cylindrical shell containing a viscous liquid is of order $O(t^{1/2})$ \citep{elbaz2014dynamics}, and is independent of the inlet signal or the nature of the external forcing, this is a linear property which stems from the heat kernel and applies to the linear interaction model of the current study (\ref{LIN_IBVL_P}). Thus, when comparing the spread rate of pressure driven viscous peeling with the propagation of a disturbance over a non-negligible fluid layer, we note that an impulse of type (\ref{ZKB_IC_BC}) is slower, $O(t^{1/5})$, than its linear counterpart for large enough time, and that otherwise faster rates can be achieved due to the dependence in inlet pressure.

\section{Effect of pre-wetting layer, self-similar analysis and numerical solution}
\begin{figure}
\centering
\includegraphics[width=1.0\textwidth]{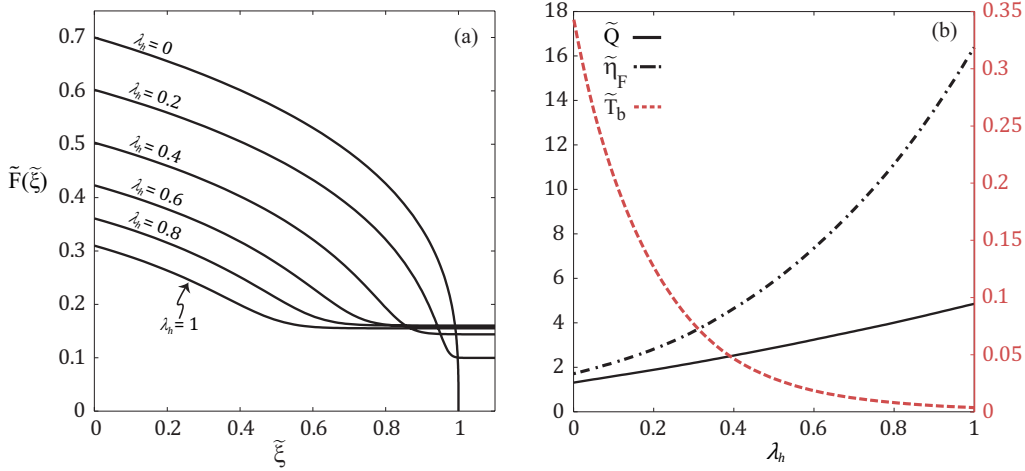}% Here is how to import EPS art
\caption{Self-similar solutions of cylindrical pressure driven viscous peeling over a pre-wetting layer for constant boundary pressure, corresponding to $0 \le \lambda_h \le1, P_e=0$ and inlet pressure $P(0,T)= 1$. Part (a): Shifted pressure profile $\widetilde{F}(\tilde{\xi})$ corresponding to eigenvalues: $\lambda _h=0,0.2,0.4,0.6,0.8,1$ (obtained numerically). Part (b): Flux rate $\widetilde{Q}(\lambda _h)$ - solid black line, front locus ${\tilde{\eta} _{_F}}(\lambda _h)$ - dashed-dotted black line, time to boundary $\widetilde{T}_b({\lambda _h})=\tilde{t}_b/\kappa t^*$ - dashed red line in right alined scale (obtained numerically).}
\label{figure_3}
\end{figure}

We consider a constant nonzero pre-wetting layer $\lambda_h > 0$, $H_0=1$, about a stationary cylindrical centre-body, with no external forcing $P_e=0$. Eq. (\ref{NL_IBVL_P}) reduces to
\begin{equation}
\frac{{\partial {P}}}{{\partial T}} - 4\frac{\partial }{{\partial Z}}\left\{ {\frac{{\partial P}}{{\partial Z}}{{\left( {{\lambda _h} + P} \right)}}^3} \right\}\sim0\,,
\label{NLDE_Pre_Wet}
\end{equation}  
%while in the range $\lambda_h \ge 1$ equation (\ref{LIN_IBVL_P}) reduces to
%\begin{equation}
%\frac{{\partial {P}}}{{\partial T}} - \frac{\partial }{{\partial Z}}\left\{ {\frac{{\partial P}}{{\partial Z}}{{\left( {1 + {{{\lambda }}_h^{-1}}{P}} \right)}}^3} \right\}\sim0\,.
%\label{LIN_Pre_Wet}
%\end{equation} 
with $T=t/\kappa t^*$ and $\kappa=48$. We transform back to PME form, $\Phi= {\lambda _h} + P$,
\begin{equation}
\frac{{\partial \Phi}}{{\partial T}} - \frac{{{\partial ^2}{{\Phi}^4}}}{{\partial {Z^2}}}\sim0{\kern 1pt}\,.
\label{PME_PHI}
\end{equation}
Focusing on the case of constant boundary pressure driven peeling,
\begin{equation}
P(0,T)=1\,,\,\,P(Z,0)=0\,,\,\,\alpha=0\,,
\label{PME_PHI_BC}
\end{equation}
the self-similar analysis of \S 4 may be extended to include the effect of the pre-wetting layer.  We adopt the tilde symbol for all self-similar variables corresponding to a pre-wetting layer. Following \S 4, we substitute $\Phi={\tilde{\eta} _{_F}}^{\frac{2}{3}}\widetilde{F}(\tilde{\xi})$, and rewrite the boundary and initial condition in shifted pressure $\Phi$ in terms of the pre-wetted self-similar pressure profile $\widetilde{F}$,
\begin{equation}
\widetilde{F}(0)={\tilde{\eta} _{_F}}^{-\frac{2}{3}}(1+{\lambda _h})\,,\,\,\,\widetilde{F}(\tilde{\xi}\ge1)={\tilde{\eta} _{_F}}^{-\frac{2}{3}}{\lambda _h}\,,
\label{Self_sim_BC_F_PW}
\end{equation}
in similar fashion to (\ref{Self_sim_BC_F}). Since we are interested in values of ${\lambda _h}$ for which $P(Z,T)$ maintains its compact support we must add the condition that,
\begin{equation}
\frac{d\widetilde{F}(1)}{d\tilde{\xi}}=0\,.
\label{Self_sim_BC_F_CS}
\end{equation} 
Integral mass conservation (\ref{Non_dim_int_MC}) becomes,
\begin{equation}
\int_0^{{\widetilde{Z}_{_F}}(T)} {P(Z,T)dZ}=\widetilde{Q}T^{\frac{1}{2}},\,\,\,\widetilde{Q}=\tilde{\eta}_{_F}^{\frac{5}{3}}\int_0^1 {\widetilde{F}(\tilde{\xi} )d\tilde{\xi} }- {\lambda _h}{\tilde{\eta} _{_F}}\,, 
\label{Non_dim_int_MC_PW}
\end{equation}
where $\widetilde{Q}$ and ${\tilde{\eta} _{_F}}$ are functions of the base layer thickness ${\lambda _h}$. The self-similar ODE ((\ref{Self_sim_ODE}), at $\alpha=0$) written in $\widetilde{F}(\tilde{\xi})$ along with boundary conditions (\ref{Self_sim_BC_F_PW}),(\ref{Self_sim_BC_F_CS}) and (\ref{Non_dim_int_MC_PW}) formulate an eigenvalue problem in ${\lambda _h}$. It is solved numerically with initial guess from \S 4 ($\alpha=0$). The solution is presented in Fig. \ref{figure_3}(a) and starts from the outer pressure profile of the unpenetrated interface (${\lambda _h}=0$) previously attained in \S 4. Subsequent profiles correspond to ${\lambda _h}=0.2,0.4,0.6,0.8$ and $1$. Fig. \ref{figure_3}(b) depicts the resulting flux rate $\widetilde{Q}$, front locus ${\tilde{\eta} _{_F}}$, and boundary time $\widetilde{T}_b$, in the nonlinear scale range $0\le {\lambda _h} \le 1$. 
We thus derive propagation laws for a constant boundary pressure viscous-elastic peeling front when spreading over a base layer. The front location ${\widetilde{Z}_{_F}}$, accumulated mass in the interface excluding the base layer $\widetilde{M}$, and time to boundary $\widetilde{T}_b$ are given by,
\begin{equation}
{\widetilde{Z}_{_F}} = {{\tilde{\eta} _{_F}}({\lambda _h})}{T^{\frac{{1}}{2}}},\,\,\,\widetilde{M}=\widetilde{Q}({\lambda _h}){T^{\frac{{1}}{2}}},\,\,\,\widetilde{T}_b={{{\tilde{\eta} _{_F}}^{-2}}({\lambda _h})}\,,
\label{Self_sim_Prop_laws_PW}
\end{equation}
where $T=t/\kappa t^*$. In the presence of a base layer, the unpenetrated interface front of \S 4 ($\alpha=0$) is accelerated by 
${\tilde \eta _{_F}}({\lambda _h})/{\tilde \eta _{_F}}(0)$, resulting in faster boundary time $\widetilde{T}_b$ and an added mass of $\widetilde{Q}({\lambda _h})/\widetilde{Q}(0)$. 

For nonzero $\lambda_h$ (\ref{NLDE_Pre_Wet}) is no longer parabolic degenerate for $P=0$ and therefore does not exhibit the orthogonal fronts of the PME (\ref{PME_P}). As ${\lambda _h}$ grows, a larger ${\tilde{\eta} _{_F}}$ is required to converge to a constant value at the right boundary ($\tilde{\xi}=1$) causing the profiles to spread inward (see Fig. \ref{figure_3}(a)). For ${\lambda _h}>1$, $\widetilde{F}$ scales as,
\begin{equation}
\widetilde{F}(\tilde{\xi};\lambda _h)=O\left(\frac{\lambda _h}{{{\tilde{\eta} _{_F}}^{\frac{2}{3}}}({\lambda _h})}\right)\,,\,\,\, {\lambda _h}>1
\label{Self_sim_F_scale}
\end{equation}
and ultimately vanishes as ${\lambda _h}\to \infty$. Thus, for ${\lambda _h}>1$ the self-similar profiles of $\widetilde{F}(\tilde{\xi})$ still provide an exact solution for $P(Z,T)$ albeit in a diminishing time frame $T<\widetilde{T}_b$, representing the limit of the nonlinear interaction model when extended into the linear ${\lambda _h}$ range. We'de like to emphasize that while constant boundary pressure driven peeling propagates as $O(t^{1/2})$ irrespective of the base layer thickness, this is not the case for other types of inlet signals. For example, source type boundary conditions (to be examined below (\ref{Pre_Wet_BC_1})) transition from $O(t^{1/5})$ propagation for ${\lambda _h} \to 0$ to $O(t^{1/2})$ for ${\lambda _h} \to \infty$.

\begin{figure}
\centering
\includegraphics[width=0.78\textwidth]{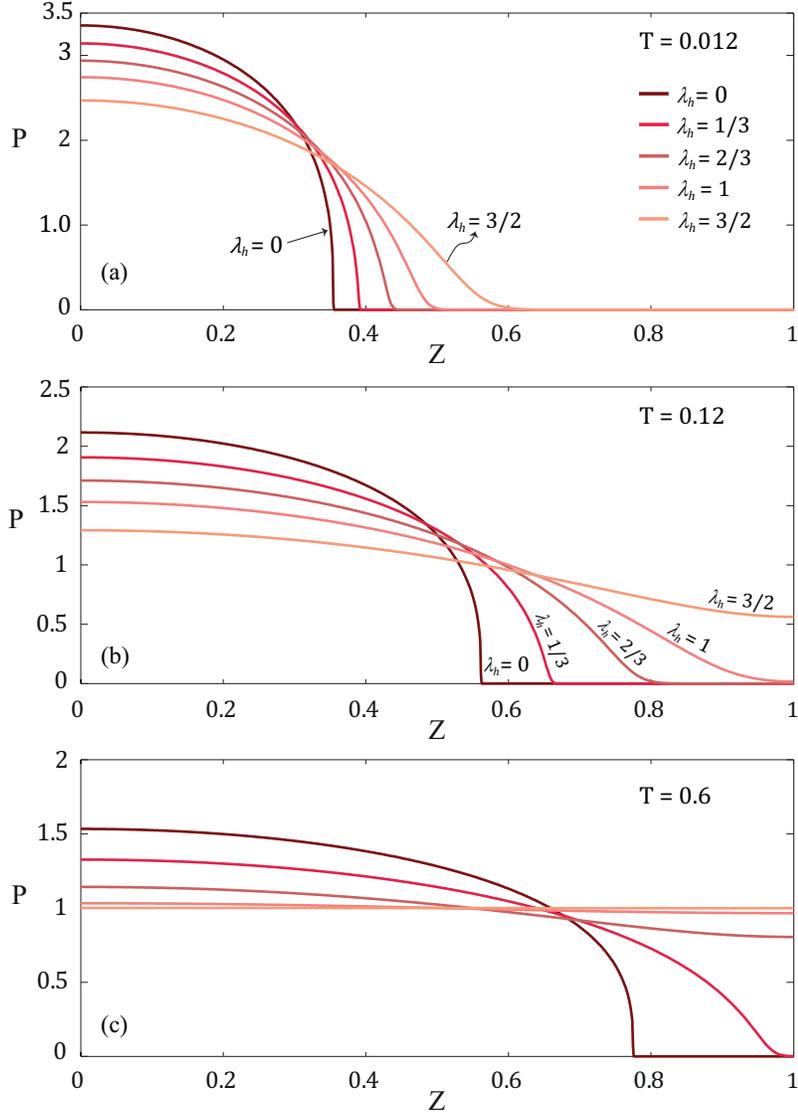}% Here is how to import EPS art
\caption{Nonlinear viscous-elastic interaction over a pre-wetting layer - source solution. Numerical solution corresponding to $0\le\lambda_h \le3/2, H_0=1, P_e=U_0=0$, $0\le T \le 0.6$. Pressure plotted vs. axial coordinate. Boundary conditions of type (\ref{Pre_Wet_BC_1}). $\lambda_h=0,1/3,2/3,1,3/2$ from dark to light reds respectively. Part (a): solution at time T=0.012, Fig. Part (b): solution at time T=0.12, Part (c): solution at time T=0.6.}
\label{figure_4}
\end{figure}

We illustrate the effect of a pre-wetting layer on the propagation of a source type viscous-elastic front via numerical solution of (\ref{NLDE_Pre_Wet}) (readjusted to the natural time-scale $T=t/t^*$) along with appropriate boundary conditions, 
\begin{equation}
P(Z,0) = \delta (Z),\,\,\,\frac{{\partial P(0,T)}}{{\partial Z}}=\frac{{\partial P(1,T)}}{{\partial Z}}=0\,.
\label{Pre_Wet_BC_1}
\end{equation}
A no flux condition at the right boundary has been added to demonstrate the interaction of the front with the opposing wall, deemed to start around $T=\kappa T_b(\alpha=1/5)$, for small enough values of $\lambda_h$. Propagation in time is shown in 3 progressive snapshots in Fig. \ref{figure_4}(a),(b),(c) corresponding to $T=0.012,0.12,0.6$, respectively, and for varying base layer thickness $0\le\lambda_h \le3/2$. The solution was obtained via a finite difference scheme and was validated on the basis of section \S4. For $\lambda_h > 1$ the solution is presented in the nonlinear time scale normalization (\ref{NL_time_scale}) for comparison. Acceleration of the front as well as its spreading is observed as the base layer thickens. At the physical limit of large ${\lambda _h}$ the solution for source type boundary conditions (\ref{Pre_Wet_BC_1}), representing pressure driven viscous-elastic perturbations of the external shell, tends to the behavior of the heat equation, $\lambda_h\to\infty$ in (\ref{LIN_IBVL_P}), which converges exponentially and for which the transfer of information is no longer confined to a compact support of the solution, which generally travels immediately to the whole interface as of $T=0$. Thus the solution of a propagating front or disturbance over a pre-wetting layer can be viewed as the synthesis of two fundamental solutions, the ZKB profiles of (\ref{ZKB_SOL}) and the Gaussian profiles akin to (\ref{LIN_IBVL_P}). As the base pressure falls below the crossover value of condition (\ref{Cnd_NL_P}) the solution will increasingly exhibit linear behavior and vice versa.  

\section{Dipole structures and their role in the transient response to moving external forces}
One of the fundamental solutions of the PME (\ref{PME_P}), as first attained by \cite{Barenblatt1957dipole}, is the dipole solution. It is constructed by taking the distributional spatial derivative of the delta function as initial condition, $P(Z,0)=\partial \delta(Z)/\partial Z$, on the infinite line. The dipole solution has also been studied in the context of viscous gravity currents \citep{king2003dipole}, analogous to the configuration presented in \S 4. However, the peeling formation of \S 4 cannot sustain negative pressure values due to violation of the surface condition (\ref{surface_cond_pres}) and therefore cannot accommodate the classical dipole solution.

\begin{figure}
\centering
\includegraphics[width=1.0\textwidth]{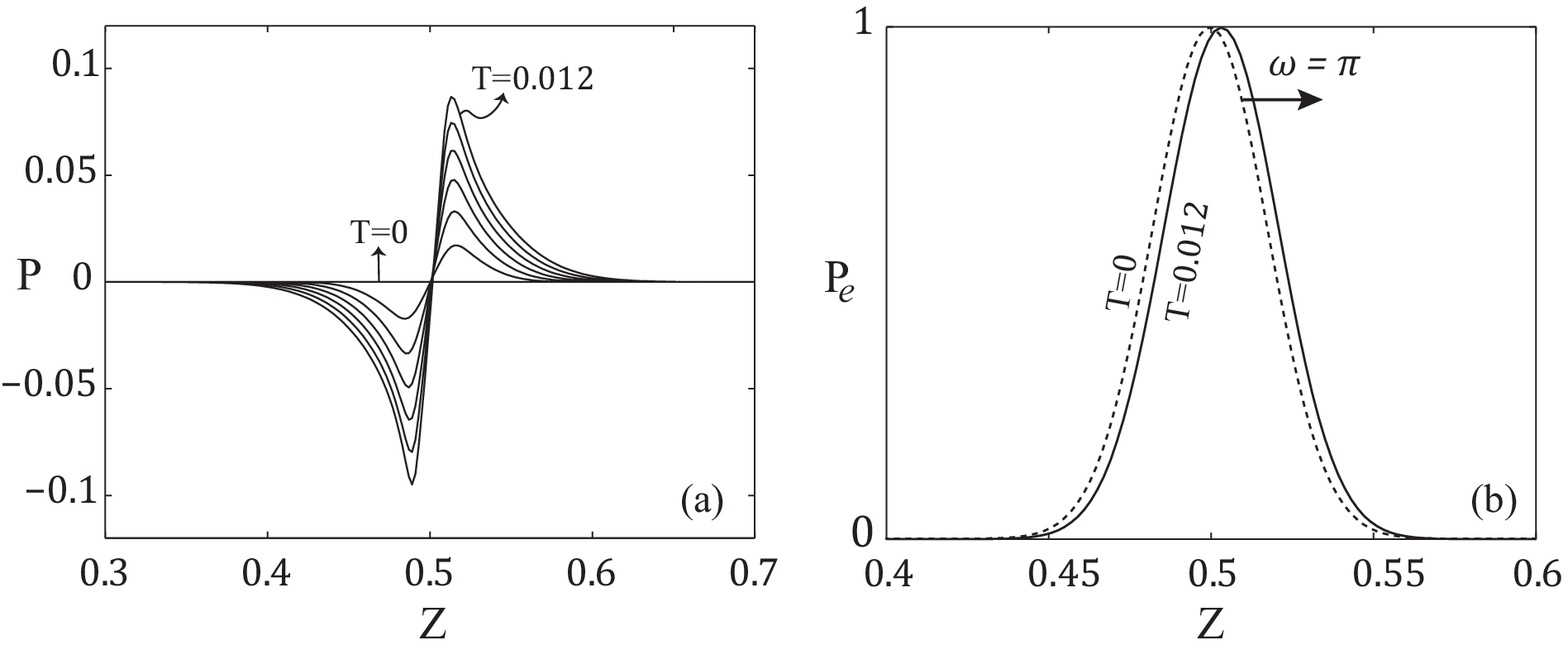}% Here is how to import EPS art
\caption{Initial transient phase dipole formation in response to an advancing spatially localized external force. Part (a): Fluid pressure dipole structure close-up, solution in $\Delta T=0.002$ increments between $0 \le T \le 0.012$. Part (b): External pressure $P_e=\delta (Z-1/2-\omega T)$, implemented as a unit Gaussian, width $0.1$, velocity $\omega=\pi$, $0 \le T \le 0.012$.}
\label{figure_6}
\end{figure}

Let us examine governing equation (\ref{NL_IBVL_P}) when forced by external pressure $P_e$. A non-negligible fluid layer resides in the annular gap $\lambda_h=1, H_0=1$ and the centre-body is stationary $U_0=0$ and of zero curvature $d{S_b}/{dZ}=0$. Hence (\ref{NL_IBVL_P}) reads,
\begin{equation}
\frac{{\partial {P}}}{{\partial T}} - \frac{1}{12}\frac{\partial }{{\partial Z}}\left\{ {\frac{{\partial P}}{{\partial Z}}{{\left( {{\lambda _h} + P-P_e} \right)}}^3} \right\}\sim \frac{{\partial {P_e}}}{{\partial T}}\,.
\label{P_dipole}
\end{equation} 
A structure similar to a dipole may occur when a high pressure region and a low pressure region form adjacent to one another while the diffusion coefficient (here $(\lambda _h + P-P_e )^3$) zeroes at the pivot (see Fig. \ref{figure_6}(a)). Such conditions can be met at the transient phase of the response to an advancing spatially localized external force,
\begin{equation}
P_e(Z,T) = \delta (Z-1/2-\omega T)\,.
\label{P_e_dipole}
\end{equation}
Fig. \ref{figure_6}(a) depicts the initial transient phase of the solution of equation (\ref{P_dipole}) under these conditions, plotted between $0\le T \le 0.012$ in $\Delta T=0.002$ increments. The forcing (\ref{P_e_dipole}) is implemented as a unit Gaussian so as to scale locally as the film layer ${\lambda _h}$, its width $0.1$  and velocity $\omega=\pi$, as depicted in Fig. \ref{figure_6}(b). Outside the dipole centre the base layer smooths out the pressure. As the external shock moves down the tube $T>0.012$ the dipole structure will collapse and the system will transition into a new pressure and deformation regime.

\begin{figure}
\centering
\includegraphics[width=1.0\textwidth]{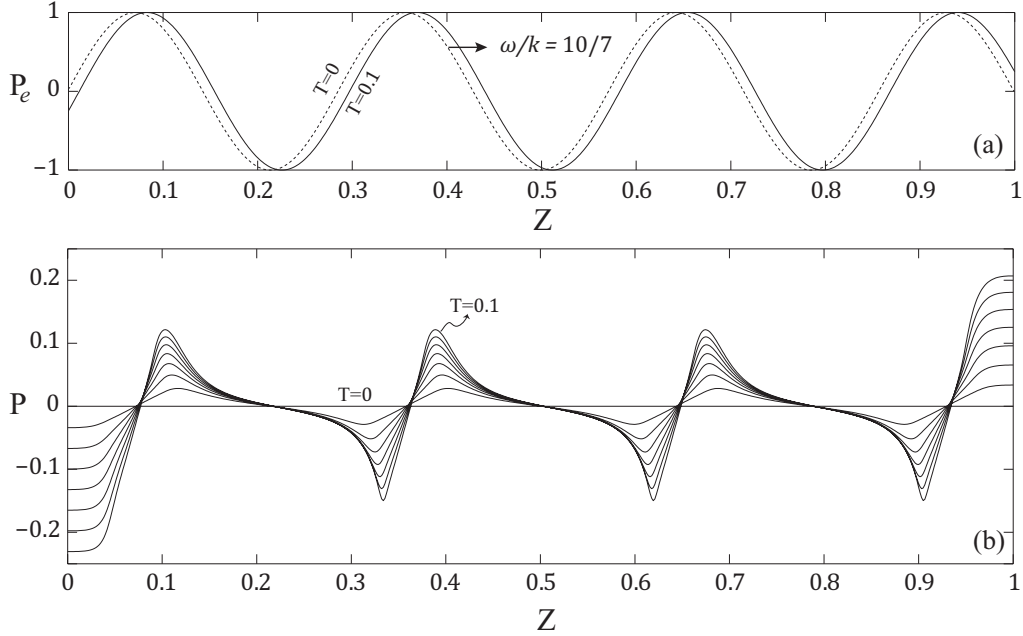}% Here is how to import EPS art
\caption{Initial transient phase dipole structure formation in response to an external moving pressure wave. Tube closed at both ends: $\partial P(0,T)/\partial Z = \partial P(1,T)/\partial Z=0$. Part (a): External pressure $P_e=a \sin(kZ-\omega T), a=1,k=7\pi,\omega=10\pi$, $0 \le T \le 0.1$. Part (b): Fluid pressure in $\Delta T=0.014$ increments between $0 \le T \le 0.1$.}
\label{figure_7}
\end{figure}

While (\ref{P_e_dipole}) is a specific case designed to create a dipole-like structure, dipoles are inherent to the
initial stages of the response to externally moving forces, occurring in areas where the diffusion coefficient vanishes. We define a general external moving wave force of the form,
\begin{equation}
P_e(Z,T) = a \sin(kZ-\omega T).
\label{P_e_dipole_osci}
\end{equation}
Fig. \ref{figure_7} demonstrates this transient dipole formation for the case of $a=1,k=7\pi,\omega=10\pi$, $P_e$ plotted in Fig. \ref{figure_7}(a) between $0 \le T \le 0.1$. A no-flux boundary condition was employed at either end of the tube,
\begin{equation}
\frac{{\partial P(0,T)}}{{\partial Z}} = \frac{{\partial P(1,T)}}{{\partial Z}}=0\,,
\label{dipole_osci_BC}
\end{equation}
representative of a closed system. The solution (see Fig. \ref{figure_7}(b)) is plotted between $0 \le T \le 0.1$ in $\Delta T=0.014$ increments. The outermost dipoles are strongly influenced by the boundary walls. The asymmetry of the central dipoles results from the simultaneity of the transient buildup and diffusion processes. Thus, the dipole structure is a fundamental mechanism of the transient response to external forcing and is essential when considering a pulsating external environment.

\section{Transported ZKB profile, a building block for creating complex deformation patterns}
\begin{figure}
\centering
\includegraphics[width=1.0\textwidth]{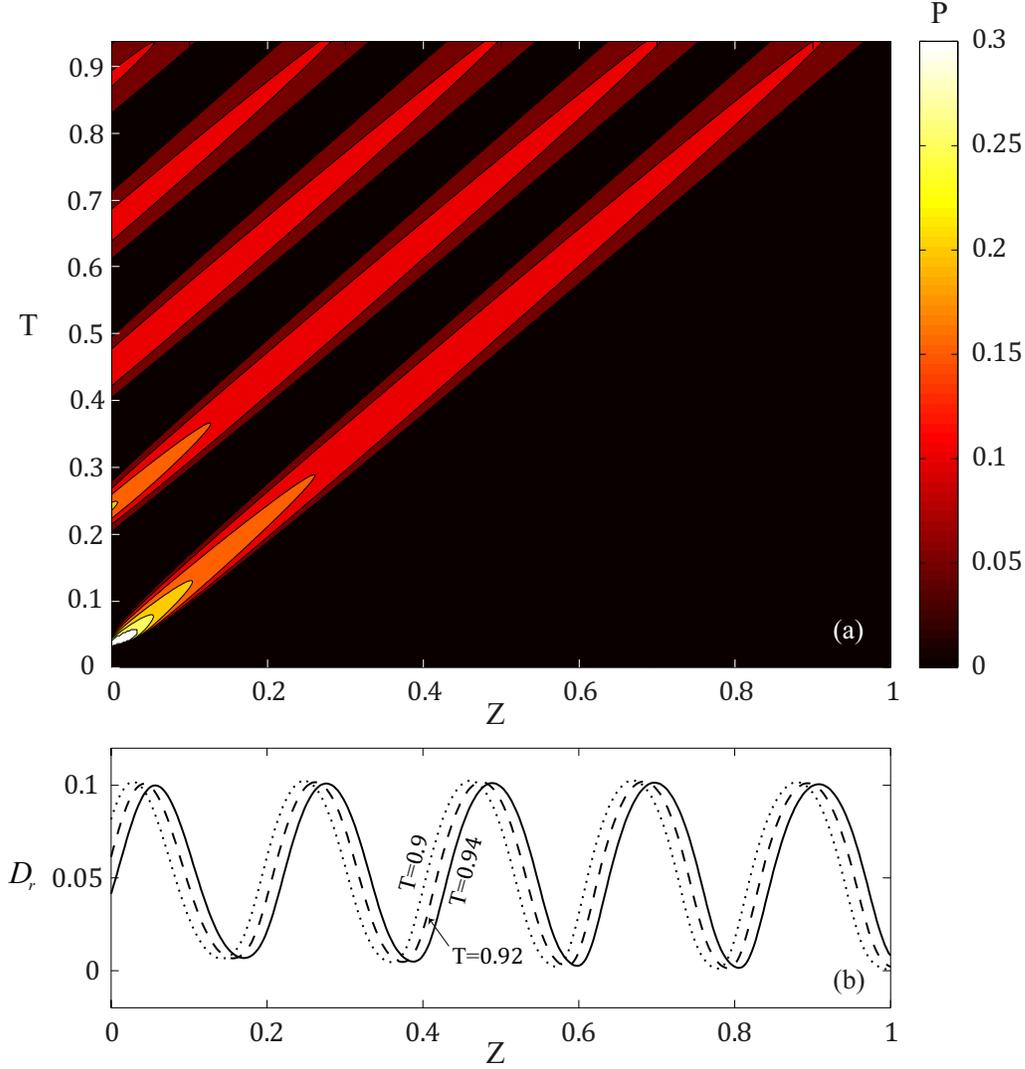}% Here is how to import EPS art
\caption{Moving wave deformation pattern generated via a modulated impulse sequence (\ref{MW_BC_1}). A pre-wetting layer of $\lambda_h=0.15$ is used upon constant centre-body motion of $U_0=2$. Modulation laws in the interval  $0\le T \le 1$: $A_n=[0.37,0.54,0.54,0.54,0.54],\tau_n=[0.003,0.02,0.0373,0.055,0.0732],{\Delta T_n}=[0.0005,0.0026,0.0036,0.0044,0.005],n=1..5$. Part (a): fluid pressure contour in space and time. Part (b): 3 progressive snapshots of the radial deformation wave at times $T=0.9,0.92,0.94$, as it propagates the interface.}
\label{figure_5}
\end{figure}

In this section we illustrate the use of the transport term of (\ref{NL_IBVL_P}) $\partial P/\partial Z\cdot U_0/2$, and the relation between the speed of pressure propagation and the gap ($\lambda_h$), in order to obtain isolated moving deformations. The linear transport term of equation (\ref{NL_IBVL_P}) may represent a relative motion of the centre-body, or alternatively, $U_0(T)$ can be thought of as a slip velocity condition induced by an electric zeta potential or a similar mechanism. We set $P_e=0$ and examine uniform geometry $H_0=1,\,\,d{S_b}/{dZ}=0$, with a small pre-wetting layer $\lambda_h\ll1$. Eq. (\ref{NL_IBVL_P}) reduces to
\begin{equation}
\frac{{\partial {P}}}{{\partial T}}+\frac{U_0}{2}\frac{{\partial P}}{{\partial Z}} - \frac{1}{12}\frac{\partial }{{\partial Z}}\left\{ {\frac{{\partial P}}{{\partial Z}}{{\left( {{\lambda _h} + P} \right)}}^3} \right\}\sim0\,.
\label{NLDE_MW}
\end{equation}  
At the left boundary we employ an impulse sequence of the form,
\begin{equation}
P(0,T) =  \sum\limits_{n = 0}^N {{A_n}\left[ {\Theta \left( {T + \frac{{\Delta T_n}}{2} - {\tau _n}} \right) - \Theta \left( {T - \frac{{\Delta T_n}}{2} - {\tau _n}} \right)} \right]} \,,
\label{MW_BC_1}
\end{equation}
where $\Theta$ is the Heaviside function. The signal is modulated in amplitude, $A_n$, and in width ${\Delta T_n}$ and can be sequenced non-uniformly by $\tau _n$. We set a zero initial condition and a no flux condition at the right boundary,
\begin{equation}
P(Z,0) = 0,\,\,\,\frac{{\partial P(1,T)}}{{\partial Z}}=0\,.
\label{MW_BC_2}
\end{equation}
Under the above conditions a sequence of symmetric ZKB profiles will propagate through the interface. Assuming sufficiently small $\lambda_h$, their compact support is maintained under transport, that is, each pulse is independent and generally does not interact with neighboring pulses. The rate of diffusion of the profiles decays strongly beneath a certain amplitude (determined by (\ref{NL_time_scale})) and is significantly slower than the rate of their transport, that is, they maintain their form as they are transported. Using this approach, isolated fluid segments can be generated arbitrarily and transported along the interface independent of the pressure gradient and having limited or negligible communication with each other. In particular, the boundary sequence (\ref{MW_BC_1}) can be modulated to form a moving wave signal, as illustrated in Fig. \ref{figure_5} via numerical solution of (\ref{NLDE_MW})-(\ref{MW_BC_2}). A pre-wetting layer of $\lambda_h=0.15$ was used with a constant centre-body motion of $U_0=2$. Fig. \ref{figure_5}(b) shows $3$ progressive snapshots of the resulting radial deformation wave at times $T=0.9,0.92,0.94$, as it propagates the interface. The corresponding fluid pressure contour is plotted in Fig. \ref{figure_5}(a) and shows the uniformity of the sequence beyond an early generation time, $T>0.35$. The obtained solution closely approximates the waveform, 
\begin{equation}
D_r(Z,T)\approx \frac{A}{2}\left[ {1 + \sin (K Z - \Omega T)} \right],\,\, \frac{\Omega}{K}=\frac{U_0}{2},
\label{MW_approx}
\end{equation}
where $A=0.1$ and $K=\Omega= 10 \pi$.

\section{Concluding remarks} 
The analysis presented here assumes several small parameters, including geometric requirements such as $\varepsilon_2=(r_o-r_i)/r_i\ll1$ and dynamic physical requirements such as small deformations $\varepsilon_3=d_r^*/r_{b0}\ll$ and negligible inertia $Re_r=\rho \epsilon^2 r_{b0}^2{p^*}^5/\mu^2 (E\varepsilon_2)^4\ll1$. While the geometric requirements are given and constant for a specific configuration, the physical requirements depend on, and limit the magnitude of, the characteristic driving pressure. To illustrate the maximal allowable pressures and characteristic time-scales we examine several configurations with constant geometric ratios $\epsilon=\varepsilon_2=0.1$. For water ($\mu=10^{-3}[Pa\cdot s]$, $\rho= 10^3 [Kg/m^3]$) as the liquid and rubber ($E=10^9[Pa]$) as the tube, we obtain $p^*=10^4[Pa], t^*=10^{4}[s]$ in the case of $r_{b0}=1[m]$ and $p^*=6.3\cdot 10^4[Pa], t^*=4[s]$ in the case of $r_{b0}=10^{-2}[m]$. For silicon oil ($\mu=10 [Pa\cdot s]$, $\rho= 7.5\cdot 10^2 [Kg/m^3]$) and rubber ($E=10^9[Pa]$), we obtain $p^*=4.2\cdot 10^5[Pa], t^*=10^{2}[s]$ in the case of $r_{b0}=1[m]$ and $p^*=2.6\cdot 10^6[Pa], t^*=0.5[s]$ in the case of $r_{b0}=10^{-2}[m]$. Hence, a wide range of characteristic driving pressures and time scales can be achieved by varying the properties of the configuration. Unlike linear viscous-elastic dynamics \citep{elbaz2014dynamics}, the assumption of creeping flow in the current nonlinear problem can be achieved for any configuration, as long as $p^*$ and $\lambda_h$ are sufficiently small.

Future research may focus on pressure and deformation propagation in cases with varying pre-wetting layer thickness $H_0(Z)$ and centre-body contour $S_b(Z)$. While the governing equation for such configurations is presented in this work (\ref{NL_IBVL_P}), the current study did not examine solutions for varying $S_b(Z)$ and $H_0(Z)$, these functions may allow for more elaborate control of the deformation field of the external shell. In addition, future research may include the effect of varying external shell thickness, which is similar to spatially varying surface tension, and may allow for transport of isolated deformations without external mechanisms such as presented in \S 7.

\appendix
% \section*{}
%\section{Error estimates due to neglected curvatures} \label{AppA}
%The neglected curvatures resulting from the time-dependent viscous elastic wave are a function of $(\partial {d_r}/\partial z)\sim{\varepsilon _1}$ and are generally of the same order-of-magnitude as the base geometrical curvatures assumed in the analysis which are a function of $(d{r_m}/dz)\sim(d{r_b}/dz)\sim{\lambda _b}{\varepsilon _1}$ (${\lambda _b} \sim 1$ or smaller). It is assumed that the shell contour, while not identical to, generally follows the center-body contour. Hence $\lambda_h=O(\lambda_b)$ when the curvature is nonzero. The error introduced in the momentum equations due to the base curvature is of the order of the ratio of the inverse radii of curvature of the shell $O({\kappa _z}/{\kappa _\theta })$, given by
%\begin{equation}
%{\kappa _\theta } = \frac{1}{{r\sqrt {1 + r{{_b^{'}}^2}} }}\,,
%\end{equation}
%\begin{equation}
%{\kappa _z} =  - \frac{{r_b^{''}}}{{{{\left( {1 + r{{_b^{'}}^2}} \right)}^{\frac{3}{2}}}}}.
%\end{equation}
%where we've denoted $r{_b^{'}}\buildrel \Delta \over =dr_b/dz$. Thus the error may be assessed,
%\begin{equation}
%O\left( {\frac{{{\kappa _z}}}{{{\kappa _\theta }}}} \right) = {\lambda _b}{\varepsilon _1}{\epsilon}\,.
%\end{equation}

\section{Summary of results in dimensional form} \label{AppA}
Nonlinear viscous elastic interaction - governing equation in fluid pressure corresponding to (\ref{NL_IBVL_P}).
\begin{equation}
\begin{split}
\frac{{\partial (p - {p_e})}}{{\partial t}} - \frac{{{r_{b0}^2}}}{{12\mu {{(E{\varepsilon _2})}^2}}}\frac{\partial }{{\partial z}}\left\{ {\frac{{\partial p}}{{\partial z}}{{\left( {\frac{{E{\varepsilon _2}{h_0}}}{{{r_{b0}}}} + p - {p_e}} \right)}^3}} \right\}\\ + {u_0}\left[ {\frac{1}{2}\frac{\partial }{{\partial z}}\left( {\frac{{E{\varepsilon _2}{h_0}}}{{{r_{b0}}}} + p - {p_e}} \right) + \frac{{E{\varepsilon _2}}}{\epsilon}\frac{{d{s_b}}}{{dz}}} \right]\sim0\,.
\label{NL_IBVL_P_dim}
\end{split}
\end{equation}
Closed-form solution of impulse driven viscous peeling according to self-similarity, corresponding to (\ref{ZKB_SOL}).
\begin{equation}
p(z,t) = {p^*}^{\frac{2}{5}}{\left( {\frac{{\kappa \mu {{(E{\varepsilon _2})}^2}}}{{{\epsilon^2}}}} \right)}^{\frac{1}{5}}{t^{-\frac{1}{5}}}{{\left[ {1 - \frac{{3{p^*}^{\frac{6}{5}}}}{{40{l^2}}}{{\left( {\frac{{\kappa \mu {{(E{\varepsilon _2})}^2}}}{{{\epsilon^2}}}} \right)}^{\frac{2}{5}}}{z^2}{t^{-\frac{2}{5}}}} \right]}_{+}^{\frac{1}{3}}}\,.
\label{ZKB_SOL_dim}
\end{equation}
Inlet pressure driven viscous peeling - self-similar propagation laws corresponding to (\ref{Self_sim_Prop_laws}). 
\begin{subequations}
\label{eq:dim prop laws}
\begin{equation}\label{front_locus}
{z_{_F}}(t,\alpha) = l{\left( {\frac{{{p^*}^3{\epsilon^2}t}}{{\kappa \mu {{(E{\varepsilon _2})}^2}}}} \right)^{\frac{{1 - 3\alpha }}{2}}}{\eta _{_F}}(\alpha)\,,
\end{equation}
\begin{equation}
v(t,\alpha) = \frac{{2\pi {r_{{b0}}^3}}}{\epsilon}{\kappa ^{\frac{{5\alpha  - 1}}{2}}}{(E{\varepsilon _2})^{5\alpha  - 2}}{{p^*}^{\frac{{5 - 15\alpha }}{2}}}{\left( {\frac{t}{\mu }} \right)^{\frac{{1 - 5\alpha }}{2}}}Q(\alpha )\,,
\end{equation}
\begin{equation}
{t_b}(\alpha) = \frac{{\kappa \mu {{(E{\varepsilon _2})}^2}}}{{{p^*}^3{\epsilon^2}}}{\eta _{_F}}{(\alpha )^{ - \frac{2}{{1 - 3\alpha }}}}\,.
\end{equation}
\end{subequations}
$v$ is the interface volume.

\acknowledgments{This research was supported by the ISRAEL SCIENCE FOUNDATION (Grant No. 818/13).}

\end{document}